\documentclass[prl,twocolumn,floatfix,showpacs,superscriptaddress]{revtex4}

\usepackage{amsmath}
\usepackage{amsfonts}
\usepackage{graphicx}

\begin{document}

\title{Joule expansion of a pure many-body state}
\author{S. Camalet}
\affiliation{Laboratoire de Physique Th\'eorique de la Mati\`ere Condens\'ee, UMR 7600, 
Universit\'e Pierre et Marie Curie, Jussieu, Paris-75005, France}
\date{Received: date / Revised version: date }
\begin{abstract}
We derive the Joule expansion of an isolated perfect gas from the principles of quantum mechanics. 
Contrary to most studies of irreversible processes which consider composite systems, the gas many-body 
Hilbert space cannot be factorised into Hilbert spaces corresponding to interesting and ignored degrees 
of freedom. Moreover, the expansion of the gas into the entire accessible volume is obtained for pure 
states. Still, the number particle density is characterised by a chemical potential and a temperature. 
We discuss the special case of a boson gas below the Bose condensation temperature.
 
 \end{abstract} 


\maketitle

Irreversible processes can be described within the framework of quantum mechanics by taking into 
account a large amount of degrees of freedom. A small number of these degrees of freedom are 
involved in the considered process and the other ones are ignored. Most of the studies are concerned 
with composite systems consisting of a subsystem of interest coupled to one or several large reservoirs. 
For example, the decoherence and relaxation into thermal equilibrium of a quantum system is obtained 
by coupling it to a large heat bath \cite{QDS}. In the context of transport, the quantum conductor under 
study is maintained out-of-equilibrium by connecting it to large free particle reservoirs \cite{Datta}. 
The interesting physics occurs generally in the subsystem and the reservoirs degrees of freedom are 
traced out \cite{Augsburg}. The reservoirs are assumed large enough so that their states remain 
essentially unchanged during the studied process. It seems then natural to suppose that they are initially 
at thermal equilibrium. 

It is often argued that the equilibrium states of the reservoirs result from their coupling to a thermal 
super-reservoir which is not taken into account explicitly in the model. Recently, it has been shown that 
such an ad hoc assumption is actually not necessary to understand the relaxation of a quantum system 
into thermal equilibrium \cite{Tasaki, Mahler,CT,EPJB}. It has been found that a boson bath initially in 
a {\it pure} state of macroscopically well-defined energy can induce this relaxation \cite{EPJB}. 
The thermalisation process has thus been obtained as a consequence of the pure quantum-mechanical 
description of a truly isolated composite system. In this approach, the temperature of the asymptotic 
equilibrium state of the interesting subsystem is not introduced by hand by thermal averaging over 
the bath initial states. It is determined by the density of states of the bath and the macroscopic energy 
of its initial pure state. 

In this Letter, we consider an isolated perfect quantum gas confined in a finite region of space. The gas 
particles do not interact with each other or with environmental degrees of freedom, they are only subject 
to a static confining potential. We show that the textbook example of an irreversible process, the Joule 
expansion, is a direct consequence of the quantum mechanics principles. More precisely, we study the time 
evolution of the number particle density ensuing from an initial {\it pure} many-body state situated in 
a subregion of the total accessible volume. Following Refs. \cite{Mahler,CT,EPJB}, we consider pure states 
of macroscopically well-defined energy. Contrary to the usually studied case of composite systems, 
the interesting and ignored degrees of freedom are here part of the same many-body system. Consequently, 
the duration of the investigated process increases with the system size and we must study a finite system. 
However, we obtain, in the thermodynamic limit, a clear expansion of the gas into the entire accessible region 
for times much shorter than the Poincar\'e recurrence time of the system. 

To simplify, we first restrict ourselves to the case of a one-dimensional perfect gas confined in a box of 
length $L$. The $N$ indistinguishable particles, bosons or fermions, of mass $m$ constituting the gas are 
described by the Hamiltonian  
\begin{equation}
H_L =-\frac{1}{2m} \int_0^L dx \psi^{\dag} (x) \partial_x^2 \psi
=\sum_{k>0}  \frac{k^2}{2m} c^\dag_k c^{\phantom{\dag}}_k \label{H}
\end{equation}
where $\psi^\dag (x)$ and $c^\dag_k$ create, respectively, a particle at position $x$ and in the single-particle 
eigenstate $k$. The sum runs over the wavevectors $k=n\pi/L$ where $n$ is a positive integer. We use units 
in which $\hbar=k_B=1$. In the following, we study the particle number density of the gas which can be written as 
\begin{equation}   
\rho (x,t)=\sum_k e^{ikx} \langle {\hat \rho}_k (t) \rangle \label{dens}
\end{equation}
where the sum runs over both positive and negative wavevectors $k$ and $\langle \ldots \rangle$ denotes 
the average with respect to the initial gas state $|\psi \rangle$. The operators ${\hat \rho}_k$ are given by
\begin{equation}   
{\hat \rho}_k (t)= \frac{1}{2L}\sum_{k' \ne 0,k} e^{ik(2k'-k)t/2m}
{\tilde c}^\dag_{k'} {\tilde c}^{\phantom{\dag}}_{k'-k}  
\end{equation}
where ${\tilde c}_k=\mathrm{sgn} (k) c_{|k|}$. We remark that 
${\hat \rho}_k^\dag={\hat \rho}_{-k}^{\phantom{\dag}}={\hat \rho}_k^{\phantom{\dag}}$ 
and $\langle {\hat \rho}_0 (t) \rangle=N/L$ for any state $|\psi \rangle$. 

\begin{figure}
\centering \includegraphics[width=0.45\textwidth]{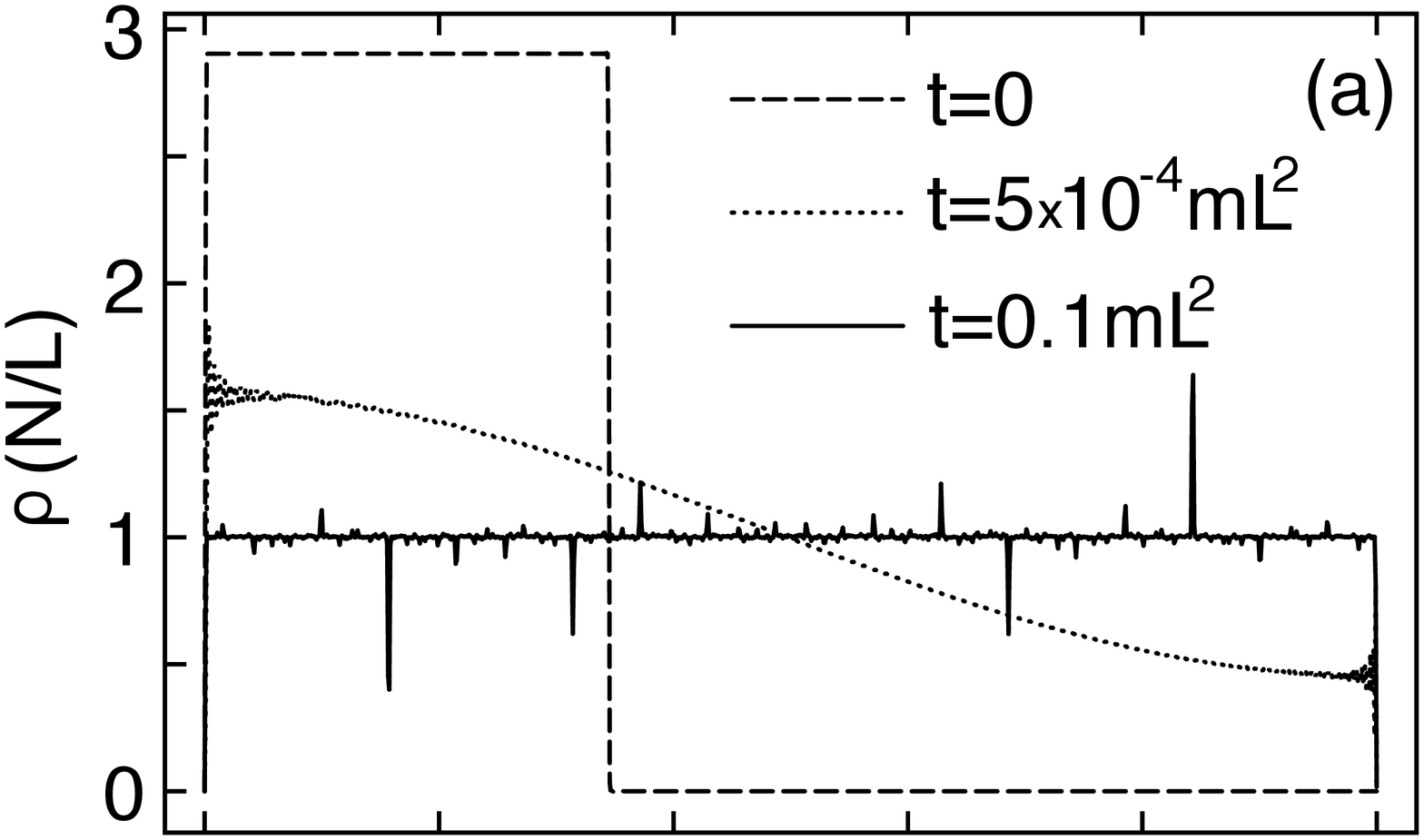}
\centering \includegraphics[width=0.45\textwidth]{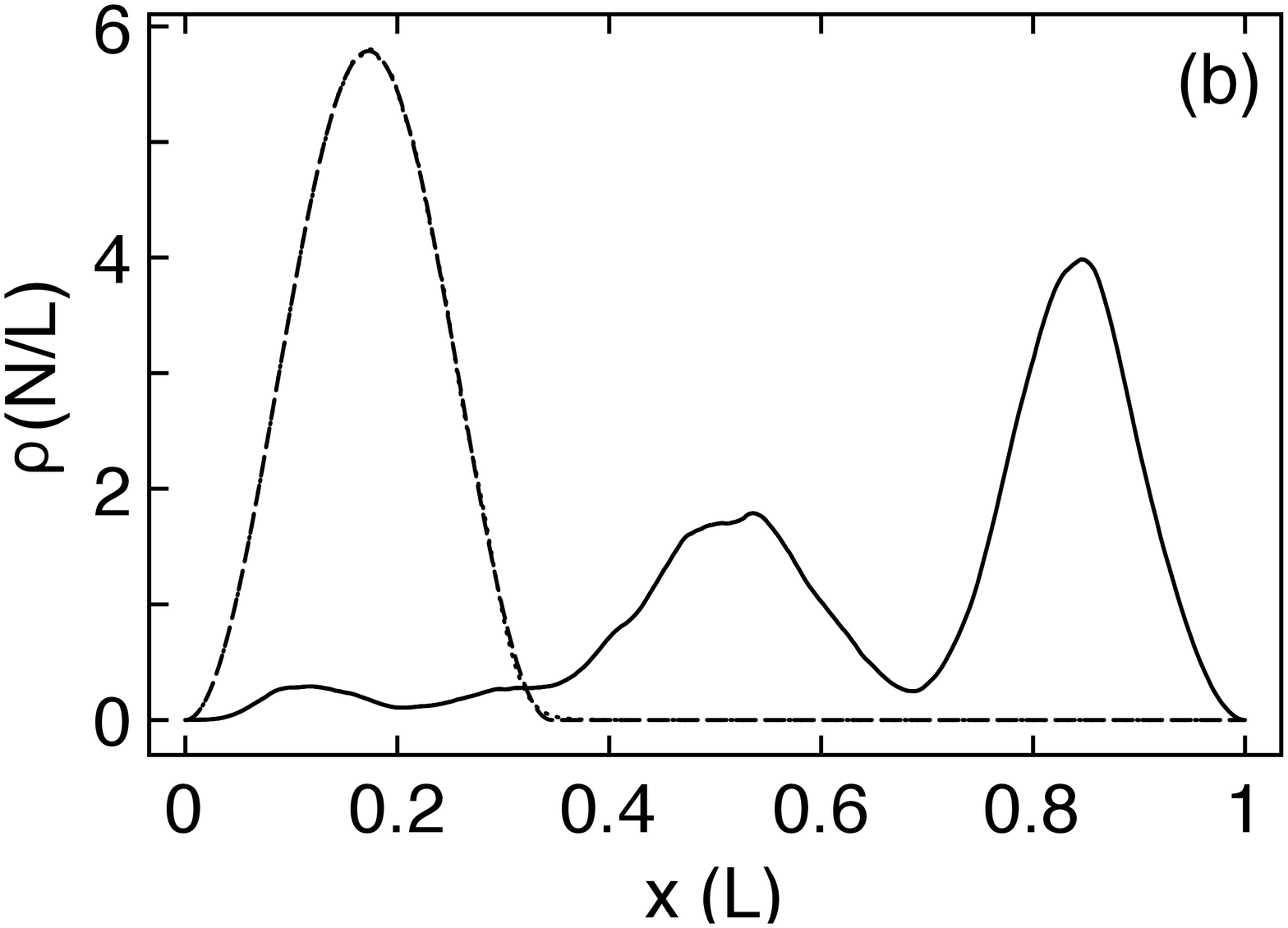}
\caption{\label{fig:rho} Particle number density as a function of space for $\ell/L=0.34567$ and 
times $t/mL^2=0$, $5 \times 10^{-4}$ and $0.1$ for (a) a 1D fermion gas at $T=10^5/m\ell^2$ and 
$\exp(-\mu/T)=0.1$ (b) a 3D boson gas at $T=0$ or, equivalently, a single particle. In this last case, 
the first two curves cannot be distinguised.  
}
\end{figure}

We assume that initially the $N$ particles are confined in an interval $[0,\ell]$ where $\ell<L$, and 
that $|\psi \rangle$ is a state of macroscopically well-defined energy $E$. We are interested in 
the subsequent evolution of the density \eqref{dens} in the thermodynamic limit $N \gg 1$ with finite 
mean distance $\ell/N$ and energy $E/N$. The initial gas state reads 
\begin{equation}
|\psi \rangle = \sum_{|\alpha\rangle \in {\cal H}_E} \psi_\alpha |\alpha \rangle \label{psi}
\end{equation} 
where $|\alpha \rangle$ refers to the $N$-particle eigenstates of the Hamiltonian $H_\ell$ given by 
\eqref{H} with the upper limit $L$ replaced by $\ell$. The wavevectors of the corresponding eigenmodes 
are denoted by $q=n\pi/\ell$ where $n$ is a positive integer. An eigenstate $|\alpha \rangle$ corresponds 
to a set of occupation numbers $\{ n_q \}$ obeying $\sum_{q>0} n_q=N$. For fermions, $n_q$ is restricted 
to the values $0$ and $1$. The Hilbert space ${\cal H}_E$ is spanned by the states $|\alpha \rangle$ 
satisfying $E<\sum_{q>0} n_q q^2/2m <E+\delta E$ where $\delta E$ is much smaller than $E$ but much 
larger than the maximum level spacing of $H_\ell$. In the thermodynamic limit, the dimension $D$ of 
${\cal H}_E$ is practically proportionnal to $\delta E$ and a density of states $n (E,N)=D/\delta E$ can be 
defined. Moreover, this density satisfies the Boltzmann's relation
\begin{equation}
\ln \left[ n (E,N) \delta E \right]  \simeq N s\left( \frac{E}{N}, \frac{\ell}{N} \right)  \label{n}
\end{equation}  
where $s=S/N$ is the gas entropy per particle. As is well known, for a perfect gas, the relation \eqref{n} 
is a direct consequence of the exchange symmetry principle  \cite{Diu}.  

We now show that, in the thermodynamic limit, almost all normalised states \eqref{psi} lead to the same 
particle number density. To obtain this result, we use, following Refs. \cite{Mahler,CT,EPJB}, the uniform 
measure on the unit sphere in ${\cal H}_E$
\begin{equation}
\mu \big( \{\psi_\alpha\} \big) = \frac{(D-1)!}{\pi^D} \delta 
\Big( 1-\sum_{|\alpha\rangle \in {\cal H}_E} |\psi_\alpha|^2 \Big). \label{mu}
\end{equation}
More precisely, we will evaluate the Hilbert space average and variance of $\langle  {\hat \rho}_k (t) \rangle$ 
following from this normalised distribution. To do so, we note that, in the limit $N \gg 1$, the reduced distribution 
of a finite number $q$ of components $\psi_{\alpha(1)}, \ldots, \psi_{\alpha(q)}$, equals 
$\prod_{p=1}^{q} D\exp(-D | \psi_{\alpha(p)} |^2)/\pi$. With this Gaussian distribution, we obtain the Hilbert 
space average
\begin{equation}   
\overline{\langle  {\hat \rho}_k (t) \rangle}  
= \frac{e^{i\epsilon_k t}}{2LD} \sum_{|\{ n_q \}\rangle \in {\cal H}_E } 
\sum_{k' ; q>0} e^{-ikk't/m} (q|k') (q|k'-k) n_q  \label{rhokmicro}
\end{equation}
where $(q|k) = 2(L\ell)^{-1/2} \int_0^\ell dx \sin(qx) \sin(kx)$ and $\epsilon_k=k^2/2m$. This expression 
simply states that the Hilbert space average of the expectation value $\langle  {\hat \rho}_k \rangle$ is equal 
to the microcanonical average at energy $E$ of the operator ${\hat \rho}_k$. It can be further simplified using 
the following standard arguments \cite{Diu}. The microcanonical probability distribution of the occupation 
number $n_q$ is $P(n_q)={\hat n}(E-n_q \epsilon_q , N - n_q )/n(E,N)$ where $\epsilon_q=q^2/2m$ and 
${\hat n}$ is the density of states of the Hamiltonian $H_\ell-\epsilon_q c^{\dag}_q c^{\phantom{\dag}}_q$. 
This density obeys the Boltzmann's relation \eqref{n} with the corresponding entropy ${\hat s}$. Expanding 
this entropy in the energy $n_q \epsilon_q \ll E$ and number $n_q \ll N$ and taking into account that 
${\hat s}=s$ in the thermodynamic limit, result in $P(n_q) \propto \exp[-n_q(\epsilon_q-\mu) /T]$ where 
$T$ and $\mu$ are given by 
\begin{equation}
\frac{1}{T} = \partial_E S (E,N) , \quad  \frac{\mu}{T} = - \partial_N S (E,N) .  \label{Tmu} 
\end{equation}
Consequently, the microcanonical average numbers in \eqref{rhokmicro} can be replaced by the Bose-Eintein 
or Fermi-Dirac occupation function at the microcanonical temperature $T$ and chemical potential $\mu$ 
determined by the intensive parameters $E/N$ and $\ell/N$ of the many-body state \eqref{psi}. 

We define the Hilbert space variance $\sigma^2$ of $\langle  {\hat \rho}_k  \rangle$ as the average 
of $\big|\langle {\hat \rho}_k \rangle - \overline{ \langle {\hat \rho}_k \rangle} \big|^2$ with respect to 
the measure \eqref{mu}. We find 
\begin{equation}
\sigma^2=\frac{1}{D^2}\sum_{|\alpha\rangle,|\beta\rangle \in {\cal H}_E} 
|\langle \alpha | \rho_x (t) |\beta \rangle |^2 
< \frac{1}{D} \langle {\hat \rho}_k (t)^2 \rangle_E \label{var}
\end{equation} 
where $\langle A \rangle_E=\sum_{|\alpha \rangle \in {\cal H}_E} \langle \alpha | A | \alpha \rangle/D$ 
denotes the microcanonical average at energy $E$ of the observable $A$. The upper bound is simply 
obtained by replacing the sum over the states $|\beta\rangle \in {\cal H}_E$ by a sum over all the $N$-particle 
eigenstates $|\beta\rangle$ of the Hamiltonian $H_\ell$. From the above arguments, one finds that 
the microcanonical distribution of two eigenmode occupation numbers is $P(n_q)P(n_{q'})$ with $T$ 
and $\mu$ given by \eqref{Tmu}. As a consequence, the microcanonical average on the right side of 
the inequality \eqref{var} is equal to a finite grand-canonical average and hence, as $D \sim \exp(N)$, 
the variance $\sigma^2$ vanishes exponentially in the thermodynamic limit. Therefore, for almost all states 
\eqref{psi}, the particle number density is given by \eqref{dens} with the Fourier coefficients
\begin{equation}   
\langle  {\hat \rho}_k (t) \rangle  
= \frac{e^{i\epsilon_k t}}{2L}\sum_{k' ; q>0} (q|k') (q|k'-k)  f (q) e^{-ikk't/m}  \label{rhokgrand}
\end{equation}
where $f (q)=[\exp(q^2/2mT-\mu/T) \mp 1]^{-1}$ depending on the bosonic or fermionic nature of 
the particles. To describe a gas state of macroscopic energy $E$, other choices than \eqref{psi} are possible. 
For example, due to the exponential $N$-dependence \eqref{n} of the density $n(E,N)$, one obtains 
the result \eqref{rhokgrand} for almost all states of the Hilbert space spanned by the eigenstates 
$|\{ n_q \}\rangle$ satisfying $\sum_q n_q=N$ and $\sum_q n_q \epsilon_q < E$. We also remark that 
the above derivation is not restricted to the observables ${\hat \rho}_k$. The result \eqref{rhokgrand} can 
be generalised to any $n$-particle observables for $n \ll N$. The expression \eqref{rhokgrand} holds for 
higher-dimensional systems, provided that the confining potential is independent of $x \in [0,L]$. More 
precisely, the number of particles $\rho(x)dx$ between $x$ and $x+dx$ evolves according to 
\eqref{rhokgrand} with $f$ replaced by a sum over transverse eigenmodes. We consider the case 
of macroscopic transverse dimensions towards the end of the paper.

\begin{figure}
\centering \includegraphics[width=0.45\textwidth]{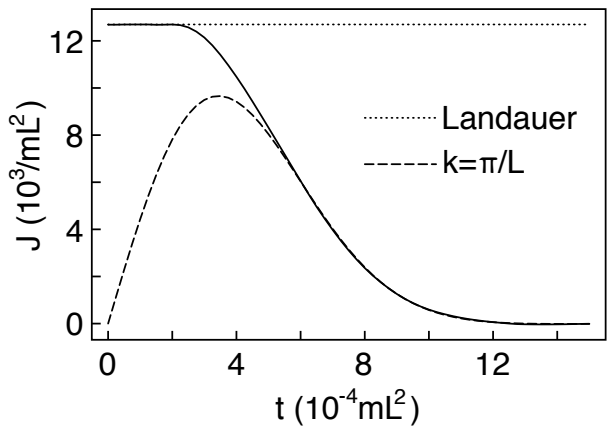}
\caption{\label{fig:current} Particle current $J$ as a function of time for fermions, a fugacity of $0.1$, 
$T=10^5/m\ell^2$ and $\ell=0.34567 L$. The dotted and dashed lines are, respectively, 
the Landauer current and the contribution of the wavevector $k=\pi/L$. }
\end{figure}

We now discuss the time evolution of the density $\rho$. For that purpose, it is useful, using the properties 
of the Dirac comb function \cite{bookFT}, to rewrite \eqref{rhokgrand} as
\begin{eqnarray}   
\langle  {\hat \rho}_k (t) \rangle  
&=& \frac{e^{i\epsilon_k t}}{4\pi L}\sum_{n,p}  \int_{-\ell}^\ell dx  \int_{-\ell}^\ell dx'  
\delta \big( x-x'+\frac{kt}{m} + 2nL \big) \nonumber \\
&~& \times  e^{ikx} \sum_{\eta=\pm 1} \eta F(x-\eta x'+2p\ell) \label{rhokfinal}
\end{eqnarray}
where $F(x) = \int dq e^{iqx} f(q)$ is the Fourier transform of the grand canonical occupation function $f$.
The function $F$ does not depend explicitly on the length $\ell$, it depends only on the intensive parameters 
$E/N$ and $\ell/N$ via the definitions \eqref{Tmu}. It assumes its maximum value $2\pi N/\ell$ at $x=0$ and 
vanishes for large $x$. As an example, consider the classical Maxwell-Boltzmann limit, $\Lambda \ll \ell/N$ where 
$\Lambda=(\pi N/E m)^{1/2}$ is the de Broglie thermal wavelength ($E = NT/2$ here), where 
$F(x) \propto \exp(-\pi x^2/\Lambda^2)$. The expression \eqref{rhokfinal} clearly shows that 
$\langle  {\hat \rho}_k \rangle$ is periodic with period $t_k=4Lm/k$ and 
$\langle  {\hat \rho}_k (t_k/2) \rangle= \pm \langle  {\hat \rho}_k (0) \rangle$ depending on the parity of $kL/\pi$. 
Consequently, the density $\rho$ is periodic with period $4mL^2/\pi$ and $\rho(x,2mL^2/\pi)=\rho(L-x,0)$. 
Times of the order of $mL^2$ are practically inaccessible. For example, for He atoms and a length 
$L \simeq 10$ cm, the time period of the particle number density is of the order of a few days. 

The gas expansion is described by the relaxation, for times $t \ll mL^2$, of the Fourier coefficients 
$\langle  {\hat \rho}_k (t) \rangle$ corresponding to wavelengths of the order of the box length $L$. 
For these times and wavelengths, we obtain
\begin{equation}   
\langle  {\hat \rho}_k (t) \rangle \simeq \frac{\sin (k\ell)}{2\pi k L} F \left( \frac{kt}{m} \right), \label{rhokapprox}
 \end{equation}
up to a correction of order $L^{-1}$. The macroscopic wavelength components of the density $\rho$ 
are of order unity at initial time and then relax according to \eqref{rhokapprox}. The gas evolution is hence 
essentially an expansion into the entire accessible volume, as illustrated by Fig.\ref{fig:rho}. This figure 
clearly shows that this evolution is very different from that of a single particle \cite{Aslangul}. The results 
displayed in Figs.\ref{fig:rho} and \ref{fig:current} are obtained by numerical evaluation of the sum 
\eqref{rhokgrand}. We remark that in the Maxwell-Boltzmann limit, the gas expansion described by 
\eqref{rhokapprox} is characterised by an $\hbar$-independent time of the order of $L(m/k_B T)^{1/2}$ 
as expected from classical physics dimensional considerations. In the example already mentionned of 
He atoms in a box of $10$ cm length, this time is of the order of $0.1$ ms  at $T=300$ K and is thus far 
much shorter than $mL^2$.  

Interestingly, the macroscopic wavelength approximation \eqref{rhokapprox} can be interpreted in 
quasiclassical terms as follows. The number particle density resulting from this approximation can be written 
as an integral over $p$ of the distribution $w(x,p,t)= f(p) \sum_n \Pi_\ell ( x-pt/m+2nL )/2\pi$ where 
$ \Pi_\ell ( x)=1$ for $|x|<\ell$ and $0$ otherwise. This probability density describes classical particles which 
are initially distributed uniformly in the sub-box $[0,\ell]$ with momenta $p$ distributed according to 
the grand-canonical occupation function and which evolve freely between perfectly elastic collisions with 
the walls of the container $[0,L]$ \cite{class}. The gas expansion given by \eqref{rhokapprox} can be 
characterised by the time evolution of the particle current $J(t)=-\partial_t \int_0^\ell dx \rho(x,t)$. For short 
times, $J$ is constant and equal to the Landauer current \cite{Landauer} 
$J_L = \int_{-\mu}^{\infty} d\epsilon /2\pi[\exp(\epsilon/T) \mp 1]$
flowing through a perfect conductor from a reservoir at thermal equilibrium, the interval $[0,\ell]$ here, to 
an empty reservoir, the interval $[\ell,L]$ here, see Fig \ref{fig:current}. For longer times, the density $\rho$ 
and hence the current $J$ are well approximated by the longest wavelength terms of \eqref{dens}, see 
Figs. \ref{fig:rho} and \ref{fig:current}. Finally, the current $J$ essentially vanishes.
 
Contrary to the quasiclassical approximation \eqref{rhokapprox}, the quantum number particle density 
determined by \eqref{rhokgrand} does not relax into a steady state. As $\langle  {\hat \rho}_k \rangle$ is 
time periodic with period $t_k=4Lm/k$, the density $\rho$ fluctuates constantly, even in the 
Maxwell-Boltzmann limit. For the sake of clarity, we discuss this limit and a length $L > 2\ell $. From 
\eqref{rhokfinal}, we deduce that $|\langle  {\hat \rho}_k (t) \rangle|$ is even and periodic with period 
$t_k/2$. Moreover, we find, for $0<t<t_k/4$, $|\langle  {\hat \rho}_k (t) \rangle|  \simeq  f( k/2 )/4L$ for 
$t/t_k<\ell/2L$ and $0$ elsewhere, provided that $t/t_k, |t/t_k - \ell/2L| \gg \Lambda/L$ where $\Lambda$ 
is the de Broglie thermal wavelength. To show that the microscopic fluctuations of the density $\rho(x,t)$ 
do not dissapear in the thermodynamic limit, we consider 
\begin{equation}   
M(t)=\int_0^L dx \left( \rho(x,t) - \frac{N}{L} \right)^2= L \sum_{k \ne 0} |\langle  {\hat \rho}_k (t) \rangle|^2 
\end{equation}
which is a global measure of the difference between $\rho$ and the uniform density $N/L$. The distance 
$M$ decreases from the extensive value $M(0) \simeq (L/\ell-1)N^2/L$ to finite values for $t \gg mL\Lambda$. 
For these times, all the Fourier components $|\langle  {\hat \rho}_k \rangle|$ are of the order of $L^{-1}$ and 
$M$ can be written as an integral over $k$ on the intervals $[2m(pL-\ell)/t,2m(pL+\ell)/t]$ where $p$ is an integer. 
Consequently, the function $\rho(x,t) - N/L$ is non-vanishing in finite size regions, see Fig.\ref{fig:rho}, and changes 
with a characteristic time of the order of $mL\Lambda$. 

We finally discuss the case of a gas confined in a tridimensionnal box of length $L$ and rectangular cross-section 
$S \sim L^2$. We assume that the gas is initially in a state of macroscopically well-defined energy $E$ and 
situated in a sub-volume $V=S\ell$ between $x=0$ and $x=\ell < L$. The main difference with the unidimensional 
system is that here, for bosons, the average occupation number $N_0$ of the sub-box $V$ lowest eigenmode is 
macroscopic for $E$ below the Bose condensation energy $E_B \propto N (N/V)^{2/3}/m$ \cite{Diu}. We note that, 
in this case, the microcanonical probability distribution of this occupation number is not of the grand-canonical form 
\cite{BEC}. Whereas one eigenmode can be macroscopically populated here, the result $\rho = \overline{\rho}$ 
remains valid since the dimension $D$ in \eqref{var} grows exponentially with $N$. For fermions and for bosons 
with $E>E_B$, the gas expands according to \eqref{rhokfinal} with $F$ replaced by 
$F_{3D} (x) = S \int dq q \sin(qx) f(q)/2\pi x$. For bosons with $E<E_B$, the density $\rho=\rho_{BEC}+\rho_{JE}$ 
is the sum of two contributions corresponding, respectively, to the condensed and non-condensed bosons. 
The contribution $\rho_{JE}$ is the zero chemical potential limit of the above case. The other contribution is very 
different since $\rho_{BEC}(x,t)$ equals $N_0$ times a single-particle position probability density, see 
Fig.\ref{fig:rho}. Therefore, though we consider pure states, the behavior of the boson gas is radically different 
below and above the Bose condensation temperature $T_B \propto E_B/N$ \cite{Diu}. 

In conclusion, we have obtained an irreversible evolution of physically relevant degrees of freedom of a many-body 
system without any coupling to environmental degrees of freedom. More precisely, we have shown that an isolated 
quantum perfect gas confined in a finite region of space tends to fill uniformly the total accessible volume, with 
the noticeable exception of the low energy 3D Bose gas. Moreover, whereas we consider pure gas states, 
the gas expansion is characterised by a chemical potential and a temperature and is well described by 
the Landauer formula for short times. It would be interesting to study the influence of interactions between 
the particles and with an environment. For a weak coupling to an environment, the behavior of the number particle 
density should be well described by our results for short enough times. The particle-particle interactions could 
induce a complete relaxation of the single-particle density matrix into a grand-canonical equilibrium state.     

\begin{acknowledgments}
We thank R. Chitra for helpful discussions and a careful reading of the manuscript.
\end{acknowledgments}

\end{document}